\documentclass[12pt]{article}
\usepackage{graphicx}

\usepackage{epsf}
\def\beq{\begin{equation}}
\def\eeq{\end{equation}}
\def\bea{\begin{eqnarray}}
\def\eea{\end{eqnarray}}

\def\vel{\left|}
\def\ver{\right|}
\def\nnb{\nonumber}

\def\rar{\rightarrow}
\def\nnb{\nonumber}
\def\la{\langle}
\def\ra{\rangle}
\def\ba{\begin{array}}
\def\ea{\end{array}}
\def\bea{\begin{eqnarray}}
\def\eea{\end{eqnarray}}

\def\Bgll{$B_s \rar \gamma \, \ell^+ \ell^-$}

\def\vel{\left|}
\def\ver{\right|}
\def\nnb{\nonumber}

\def\rar{\rightarrow}
\def\nnb{\nonumber}
\def\la{\langle}
\def\ra{\rangle}

\begin{document}
\title{ {\Large {\bf
The effects of non-standard Z couplings on the lepton
polarizations in  $B_s \rar \gamma \, \ell^+ \ell^-$ decays }}}
\author{ {\small G\"{u}rsevil Turan}\\
{\small  Physics Department, Middle East Technical University} \\
{\small 06531 Ankara, Turkey}\\}

\begin{titlepage}
\maketitle
\thispagestyle{empty}
\begin{abstract}
The fact that the measured $B\rightarrow \pi \, \pi \, , \, \pi \, K$ branching ratios
exhibit  puzzling patterns has received
a lot of attention in the literature and motivated the formulation of a series of new physics
scenarios with enhanced Z-penguins. In this work, we analyze the effect of such an enhancement
 on the lepton polarization asymmetries of $B_s\rightarrow \gamma
\ell^+ \ell^-$ decays by applying the results of a specific
scenario  for the solution of ``$B\to\pi K$ puzzle''.
\end{abstract}
\end{titlepage}
In the Standard Model (SM),  CP violation
originates from the the Cabibbo--Kobayashi--Maskawa (CKM) three generation quark mixing matrix
\cite{CKMmat}, and this  picture successfully explains the
observed CP violation in the Kaon sector . As for the B-sector, in the near future, more number of
experimental tests  will
be possible at the the B-factories providing stringent testing of
the SM mechanism of CP violation. In the mean time, some of the current
$B$-factory data seems to be  at variance with the SM description of
CP violation  for a number of processes.
 These processes can be grouped into two
systems: $B\rightarrow \pi \pi$ , $B\rightarrow \pi K$ and
$B\rightarrow \psi K$ , $B\rightarrow \phi K$ and   have received
a lot of attention in the literature. In particular;
\begin{itemize}
\item The BaBar and Belle collaborations have  recently reported
the  branching ratios of $B_d\to\pi^0\pi^0$ decays \cite{Babar1,Belle1}, which turn
out to be as large as six times the value given in a recent
 calculation \cite{BeNe} within QCD factorization \cite{BBNS1}, whereas
 the calculation of $B_d\to\pi^+\pi^-$ gives a
branching ratio about two times larger than the current
experimental average. On the other hand, the calculation of
$B^+\to\pi^+\pi^0$ agrees with the data quite well. As was
recently pointed out \cite{Buras1}, this "$B\rightarrow \pi\pi$
hierarchy" can be conveniently accommodated in the SM through
non-factorizable hadronic interference effects.

The CLEO, BaBar and Belle collaborations have also measured the
 ratios of charged $R_{\rm c}$ and neutral $R_{\rm n}$ branching ratios for $B\rightarrow \pi K$
 \cite{Buras2} that are affected significantly by colour-allowed
electroweak (EW) penguins, and reported a pattern of $R_{\rm c}>1$
and $R_{\rm n}<1$. As noted in \cite{Buras3}, this pattern
contradicts with the one obtained for decays
$B\rightarrow \pi K$ as the extension of the results for
$B\rightarrow \pi \pi$ decays using SU(3) symmetry. It has been
pointed out  that the  quantities  that may only get
contributions from   colour-suppressed forms  do not
show any anomalous behavior like $R_{\rm c}$ and $R_{\rm
n}$ do, so that this ``$B\to\pi K$ puzzle'' may be a manifestation
of new physics (NP) in the EW penguin sector \cite{Buras1,Buras3,Buras4}.

\item The decay $B_d\to\phi K_{\rm s}$ provides another suggestive contrast
 with the SM expectation. There is a
very good agreement between the SM value of $\sin 2
\beta_{\phi K_s}$, which  is the most precise information on CP
violation in the quark sector,  and the most well established
B-factory result determined from the decay $B\rightarrow \psi
K_s$. However, for the decay $B\rightarrow \phi K_s$, whose time
dependent CP violation is predicted to be the same as in
$B\rightarrow \psi K_s$ by the SM, experimental results seem to be
providing a contrast between $\sin 2 \beta_{\psi K_s}$ and $\sin 2
\beta_{\phi K_s}$. Since within the SM, this transition is
governed by QCD penguins \cite{London} and receives sizeable EW
penguin contributions \cite{RF94,Deshpande}, with more data, this
would suggest definite evidence for NP beyond the SM.
\end{itemize}

It has been argued that a promising class of models that can
explain the discrepancies summarized above involve flavor changing
couplings of the Z-boson to $\bar{b}s$. The possibility of NP with
the dominant $Z^0$-penguin contributions was first considered in
\cite{Buras:1998ed}--\cite{Buras6}, where correlations
between rare $K$ decays and $\varepsilon^{\prime}/\varepsilon$
were studied in model-independent analyses and also within
particular supersymmetric scenarios. It was generalized to rare
$B$ decays in \cite{Buchalla01}.  Recently, in \cite{Choudhury1},
the effects of enhanced Z penguins on
the lepton polarization asymmetries of $b\rightarrow s
\ell^+\ell^-$ have been considered.

In this work we analyze the effects of such a non-standard Z coupling
on the lepton polarization asymmetries of $B_s\rightarrow \gamma
\ell^+ \ell^-$ decays by applying the results of a specific
scenario considered in  \cite{Buras1} for the solution of ``$B\to\pi K$ puzzle''.
This basically
involves considering   the decays $B\rightarrow \pi\pi$ and
$B\rightarrow \pi K$ simultaneously within the SM and its simplest
extension in which NP enters dominantly through enhanced EW
penguins with new weak phases. Since $B\rightarrow \pi\pi$ decays
are only insignificantly
effected by EW penguins, this system can be
described as in the SM and allows the extraction of the relevant
hadronic parameters by assuming only the isospin symmetry. Using
SU(3) flavor symmetry it is then determined the hadronic
$B\rightarrow \pi K$ parameters through their $B\rightarrow
\pi\pi$ counterparts. Here there are two parameters, $C(x_t)$ with $x_t=m^2_t/M^2_W$, which is the
Z- penguin function and $\theta$, its complex phase. The
relevant EW penguin parameters for $B\rightarrow \pi K$ decays are
$q$ and $\phi$, whose SM values are  $q=0.69$ and $\phi =0$. It has been  found that
pattern of $R_c>1$ and
$R_n<1$ can not be described properly for these SM values; however, treating them as free parameters
and using the hadronic $B\rightarrow \pi K$ parameters, it is
possible to convert the experimental results for $R_c$ and $R_n$
into the values of $q$ and $\phi$,
\begin{equation}
q=1.78^{+1.24}_{-0.97}\, \, , \,
\,\phi=-(85^{+11}_{-13})^{\circ}\, \, , \label{qvefi}
\end{equation}
which describe all currently available data.

For the radiative \Bgll decay, the basic quark level process is $b
\rar  s  \ell^+ \ell^- $, which can be written as
\bea
{\cal H}_{eff}& =
& \frac{\alpha G_F}{ \sqrt{2}\, \pi} V_{tb} V_{ts}^*
 \Bigg{\{} C_9^{eff} (\bar s \gamma_\mu P_L b) \, \bar \ell \gamma^\mu \ell +
C_{10} ( \bar s \gamma_\mu P_L b) \, \bar \ell \gamma^\mu \gamma_5 \ell
- 2 C_7 \frac{m_b}{q^2} (\bar s i \sigma_{\mu \nu} q^\nu P_R b)
\bar \ell \gamma^\mu \ell
 \Bigg{\}}~, \nnb \\ && \label{Hamilton}
\eea
where $P_{L,R}=(1\mp \gamma_5)/2$ , $q$ is the momentum
transfer and $V_{ij}$'s are the corresponding elements of the CKM
matrix.
The values of the individual Wilson coefficients that appear in
the SM are listed in Table (\ref{table1}).
\begin{table}
        \begin{center}
        \begin{tabular}{|c|c|c|c|c|c|c|c|c|}
        \hline
        \multicolumn{1}{|c|}{ $C_1$}       &
        \multicolumn{1}{|c|}{ $C_2$}       &
        \multicolumn{1}{|c|}{ $C_3$}       &
        \multicolumn{1}{|c|}{ $C_4$}       &
        \multicolumn{1}{|c|}{ $C_5$}       &
        \multicolumn{1}{|c|}{ $C_6$}       &
        \multicolumn{1}{|c|}{ $C_7^{\rm eff}$}       &
        \multicolumn{1}{|c|}{ $C_9$}       &
                \multicolumn{1}{|c|}{$C_{10}$}      \\
        \hline
        $-0.248$ & $+1.107$ & $+0.011$ & $-0.026$ & $+0.007$ & $-0.031$ &
   $-0.313$ &   $+4.344$ &    $-4.624$       \\
        \hline
        \end{tabular}
        \end{center}
\caption{ Values of the SM Wilson coefficients at $\mu \sim m_b $ scale.\label{table1}}
\end{table}
As for the $C_9$,  its value is given as \cite{GSW,LD}
\begin{eqnarray}
C_9^{eff}(\mu)=C_9(\mu)+ Y(\mu)\,\, , \label{C9efftot}
\end{eqnarray}
where the function $Y(\mu)$ contains a part which arises
from the one loop contributions of the four quark
operators $O_1$,...,$O_6$ whose explicit forms can be found in
\cite{GSW,Misiak}, and also a part which estimates the long-distance contributions from
intermediate states $J/\psi ,\psi^{\prime}, ... \cite{LD,KS}$.

In the SM, the Wilson coefficient $C_{10}$ is given by
\bea
C_{10} & = & \frac{-1}{\sin^2\theta_w}\, Y \, ,
\eea
with
\bea Y & = & C(x_t)-B(x_t)\, ,
\eea
where explicit formulae of $C(x_t)$ and $B(x_t)$  can be found in \cite{Misiak, BurasMunz}.
For the rare B decays with $\ell^+ \ell^-$ in the final state the short distance function $Y$
can be parametrized  as
\bea Y & = & |C| \, e^{i\theta} +0.18 \, ,
\eea
and the connection between the rare decays and the $B\rightarrow \pi K$
system is established by relating  the parameters $(C, \theta )$ to the EW
penguin parameters $(q,\phi)$ by means of a renormalization group
analysis that yields \cite{Buras1}
\bea |C| \, e^{i\theta} = 2.35
\bar{q}e^{i\phi}-0.82 \, \, , \, \, \bar{q}=q\Bigg[
\frac{|V_{ub}/V_{cb}|}{0.086}\Bigg]\label{Cq}.
\eea
Furthermore,  $Y$ is rewritten  as
\bea
Y & = & |Y|\, e^{i\theta_Y} \, ,
\eea
where the constraint that $|Y|\leq 2.2$ follows from the BaBar and Belle
data on $B \rightarrow X_s \mu^+\mu^-$ decay \cite{Kaneko03}. The
central value for $Y$ resulting from (\ref{qvefi}) violates the
upper bound of $|Y|$, however considering only the subset of those
values of $(q,\phi)$  that satisfies $|Y|= 2.2$ gives
$\theta_Y=-(103\pm12)^{\circ}$.
Now, with the contributions from the enhanced and complex value of the $bsZ$
vertex, $C_{10}$ becomes
\bea
\label{C10New}
C_{10} & = & \frac{-2.2}{\sin^2\theta_w}\,
e^{i(103/180)\pi} \, ,
\eea
and  has a magnitude twice the SM one.

Having established the general form of the effective Hamiltonian,
the next step is to calculate the matrix element of the $B_s \rar
\gamma \, \ell^+\ell^-$ decay, which is induced by the inclusive
$b\rightarrow s \gamma \, \ell^+\ell^-$ one. Thus, the related matrix element can be found
from the $b\rightarrow s  \, \ell^+\ell^-$ decay by attaching a photon line
to any charged internal or external line. However,
contributions coming from the release of the free photon from  any charged internal line will
be suppressed by a factor of $m^2_b/M^2_W$ and can be neglected. When
a photon is released from   the initial quark lines it contributes to the so-called
"structure dependent" (SD) part of the amplitude, whereas, "internal Bremsstrahlung" (IB)
 part of the amplitude arises when a photon is radiated from one of  the
final $\ell$- leptons. Therefore, the total matrix element can be written as a sum of
the SD and the IB  contributions:
\beq
{\cal M}={\cal M}_{SD}+{\cal M}_{IB} \, ,
\eeq
with
\bea
\label{mel1}
{\cal M}_{SD} & = &  \la \gamma(k) \vel {\cal H}_{eff} \ver B(p_B) \ra \,\, , \,\,
{\cal M}_{IB} = \la 0 \vel {\cal H}_{eff} \ver B(p_B) \ra \,\, .
\eea
The structures in Eq. (\ref{mel1}) are parametrized in terms of the various form factors
$f,g,f_1,g_1$, calculated in the framework of light-cone QCD sum rules
\cite{ Eilam1,Aliev2} and in the framework of the light front
quark model  \cite{Geng2}. In addition, it has
been  proposed another  model  in \cite{Kruger} for the $B\rightarrow \gamma$ form factors
which  obey all the restrictions   obtained from the  gauge
invariance combined with the large energy effective theory.

By using the explicit expressions of the parametrizations for the above matrix elements
in terms of the form factors, which can be found  in \cite{Aliev2},
we calculate dilepton mass distribution for $B\rightarrow \gamma \, \ell^- \ell^+$ decay as
\cite{Aliev1,Gursevil}
\bea
\label{dGdxdz} \frac{d \Gamma}{ds} = \frac{\alpha^3 \,
G^2_F \, |V_{tb} V^{\ast}_{ts}|^2}{2^{10} \pi^4 }
 \, m_B \Delta~,
\eea where \bea \label{bela2} \Delta &  = & \Bigg \{ (1-s)^3 v\,
\Bigg (
 \, 4 m_B^2 r \, \mbox{\rm Re}[ A_1 B_1^\ast + A_2 B_2^\ast ]
+ \, \frac{2}{3} m_B^2 \Big( \vel A_1 \ver^2 + \vel A_2 \ver^2 +
\vel B_1 \ver^2 \nnb \\ & + & \vel B_2 \ver^2\Big) ( s - r ) \Bigg
) +  4 f_B \, m_\ell  \,   {\rm ln} [u]
\,\mbox{\rm Re}[(A_1+B_1) F^\ast] \, (1-s)^2 \nnb \\
&- &  4 f_B^2 \, \vel F \ver^2 \Bigg (2 v   \frac{s}{(1-s)} + \,
{\rm ln}[u] \Bigg[
 2 +\frac{2(2 r-1)}{(1-s)} -(1-s)\Bigg]\Bigg )\Bigg \}~.
\eea

Now, we would like to discuss the  lepton polarizations in the
rare \Bgll decays. For this, it is introduced spin projection
operator $N=(1+\gamma_5 \not\! S_i)/2$
 for $\ell^-$, where $i=L,~T,~N$ correspond to
longitudinal, transverse and normal  polarizations, respectively.
In the rest frame of $\ell^-$, the orthogonal unit vectors $S_i$
are defined as
\bea
S^{\mu}_L&\equiv&(0,\vec{e}_L)=\Bigg(0,\frac{\vec{p}_1}{|\vec{p}_1|}\Bigg)\, ,\nnb\\
S^{\mu}_N&\equiv&(0,\vec{e}_N)=\Bigg(0,\frac{\vec{k}\times\vec{p}_1}{|\vec{k}\times\vec{p}_1|}\Bigg) \, ,\nnb\\
S^{\mu}_T&\equiv&(0,\vec{e}_T)=\Bigg(0,\vec{e}_N\times
\vec{e}_L \Bigg)\,.
\eea
The longitudinal unit vector $S_L$
is boosted to the CM frame of $\ell^{+}\ell^{-}$ by Lorentz
transformation:
\bea
S^{\mu}_{L,CM}& = & \Bigg(\frac{|\vec{p}_1|}{m_\ell},
\frac{E_\ell~\vec{p}_1}{m_\ell|\vec{p}_1|}\Bigg) \, ,
\eea
while $P_T$ and $P_N$ are not changed by the boost since
they lie in the  perpendicular directions. For $i=L,~T,~N$, the
polarization asymmetries $P_{i}$ of the final $\ell^-$
lepton are defined as
\begin{eqnarray}
P_{i} (s) & = & \frac{\frac{d\Gamma}{ds} (\vec{n}=\vec{e}_i)-
\frac{d\Gamma}{ds} (\vec{n}=-\vec{e}_i)}
{\frac{d\Gamma}{ds} (\vec{n}=\vec{e}_i)+
\frac{d\Gamma}{ds} (\vec{n}=-\vec{e}_i)} \label{PL}\,.
\end{eqnarray}
After some algebra, we obtain the following expressions
for the differential polarization components of the $\ell^-$ lepton in \Bgll
decays:
\bea
P_L (s) & = & \frac{1}{6 v \Delta}\Bigg \{4 m^2_B v^3 s
(1-s)^3
( \vel A_1 \ver^2 + \vel A_2 \ver^2 -\vel B_1 \ver^2 - \vel B_2 \ver^2 )\nnb \\
& + & 24  f_B m_{\ell}(s-1)\Bigg[ (s-1)\mbox{\rm Re}[ (A_1- B_1)
F^\ast ]
\Big(v+\frac{(2 r -s)}{s} {\rm ln}[u]\Big) \nnb \\
& +& (1+s)  \mbox{\rm Re}[ (A_2+ B_2) F^\ast ]\Big(v (1-s)-\frac{2
r}{s}{\rm ln}[u]\Big) \Bigg]\Bigg \} \, ,
\eea
\bea
\label{PTmp}
P_{T} (s) &=&\frac{1}{\Delta}\Bigg\{
 \frac{(2\sqrt{r}-\sqrt{s})}{s v}  (1-s)  f_{B} m_{B}\pi \Bigg( \pm s v^2 (1+s)
\mbox{\rm Re}[(A_1-B_1) F^\ast]\nnb \\
&+& (s-1)(4 r+s)\mbox{\rm Re}[(A_2+B_2)F^\ast] \Bigg)\nnb \\
&-& \frac{\pi v}{4\sqrt{s}}(s-1)^{2}  2 m_{B}^{2} \sqrt{r} (s-1) s
\, \mbox{\rm Re}[(A_1+B_1) (A_2+B_2)^\ast]\Bigg\} \, ,
\eea
\bea
\label{PNmp}
P_{N} (s) &=&  \frac{\pi }{4\Delta} (1-s)m_{B} \Bigg\{2
m_{B}
(1-s)^2\sqrt{r s}\, v^{2}   (\mbox{\rm Im}[A_1 B_2^\ast]+\mbox{\rm Im}[A_2 B_1^\ast]) \nnb \\
&-& 4  (2\sqrt{r}-\sqrt{s}) f_{B} \Big((1+s) \mbox{\rm
Im}[(A_1+B_1) F^\ast] + (1-s)  \mbox{\rm Im}[(A_2-B_2)
F^\ast]\Big) \Bigg\} \, .\nnb \\ &&
\eea

We present now our numerical analysis about the differential polarization asymmetries
$P_L (s)$, $P_T(s)$ and $P_N(s)$ of $\ell^-$ for the $B_s \rar \gamma \ell^+ \ell^- $ decays
with $\ell =\mu , \tau $,  as well as
the averaged polarization asymmetries $<P_L>$, $<P_T>$
and $<P_N>$. The input parameters used in our numerical analysis are as follows:
\begin{eqnarray}
& & m_B =5.28 \, GeV \, , \, m_b =4.8 \, GeV \, , \,m_{\mu} =0.105 \, GeV \, , \,
m_{\tau} =1.78 \, GeV \, , \nnb \\
& & f_B=0.2 \, GeV \, , \, \, |V_{tb} V^*_{ts}|=0.045 \, \, , \, \, \alpha^{-1}=137  \, \,  ,
G_F=1.17 \times 10^{-5}\, GeV^{-2} \nnb \\
& &  \tau_{B_{s}}=1.54 \times 10^{-12} \, s \, .
\end{eqnarray}
To make some numerical predictions, we also need the explicit forms of the form factors $g,~f,~g_1$
and $f_1$. In this work we have used the values calculated in the framework of light-cone QCD sum
rules given in \cite{Aliev2}.

We also like to note a technical  point about  calculations of averaged polarization asymmetries.
The averaging procedure which we have adopted is
\beq
\langle P_i \rangle = \frac{\int^{1-\delta}_{(2m_{\ell}/m_B)^2}\, P_i(s) \, \frac{d\Gamma}{ds}ds}
{\int^{1-\delta}_{(2m_{\ell}/m_B)^2}\,  \frac{d\Gamma}{ds}ds}\label{Pidef}\, .
\eeq
We note that the part of $d\Gamma/ds$ in (\ref{dGdxdz}) which receives contribution from the
$\vel {\cal M}_{IB} \ver^2 $ term has infrared singularity due to
the emission of soft photon. To obtain a finite result from these integrations, we follow
the approach described in \cite{Aliev2} and impose a cut on the photon energy,
i.e., we require $E_{\gamma}\geq 25$ MeV, which corresponds to detect only hard photons experimentally.
This cut implies that $E_{\gamma}\geq \delta \, m_B /2$ with $\delta =0.01$.

\begin{figure}
\centering
\includegraphics[width=5in]{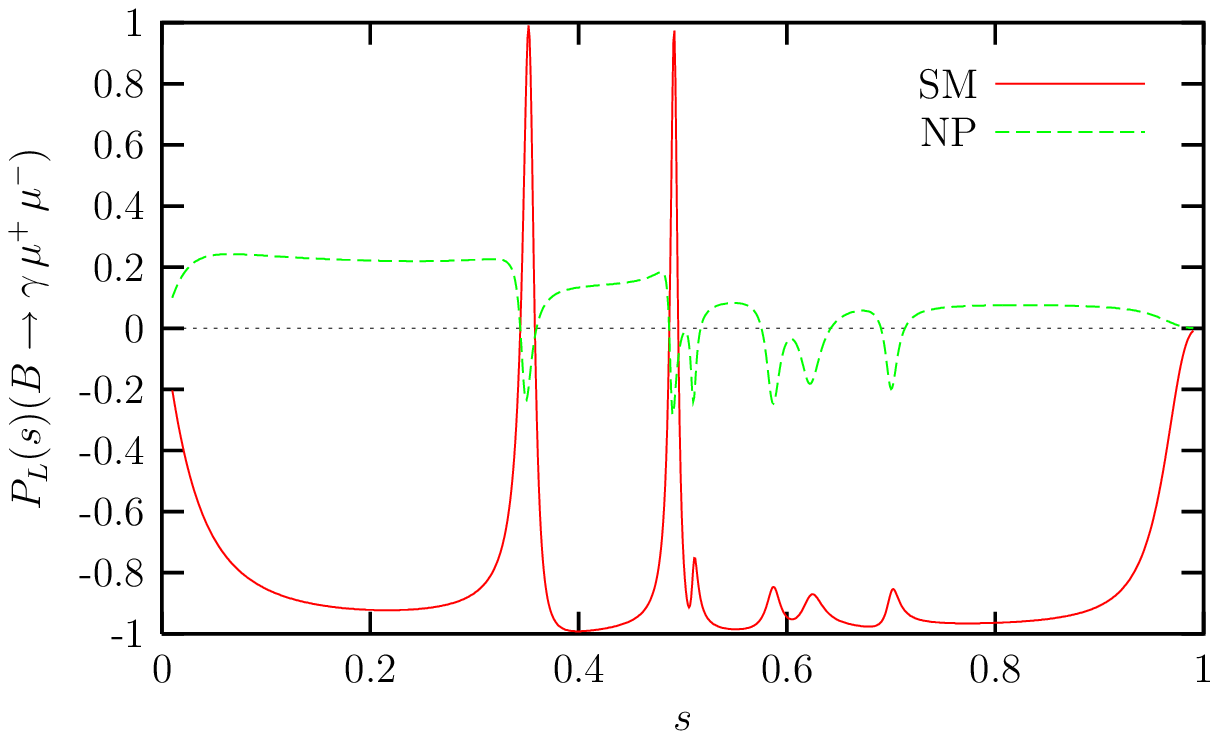}
\caption{The dependence of the  longitudinal  polarization
asymmetry $P_L (s)$ of $\ell^-$  for the $B_s \rar \gamma \, \mu^+ \mu^-$
decay on $s$ \label{f1}.}
\end{figure}

\begin{figure}
\centering
\includegraphics[width=5in]{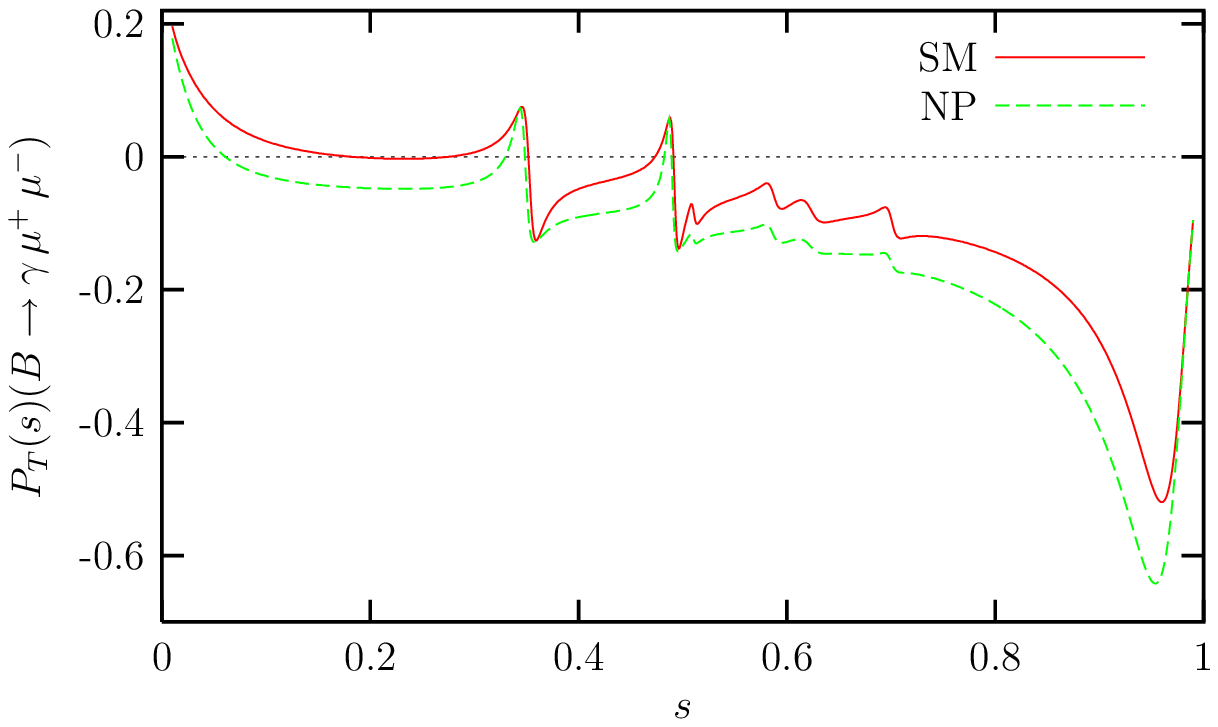}
\caption{The dependence of the  transverse  polarization
asymmetry $P_T (s)$ of $\ell^-$  for the $B_s \rar \gamma \, \mu^+ \mu^-$
decay on $s$ \label{f2}.}
\end{figure}

\begin{figure}
\centering
\includegraphics[width=5in]{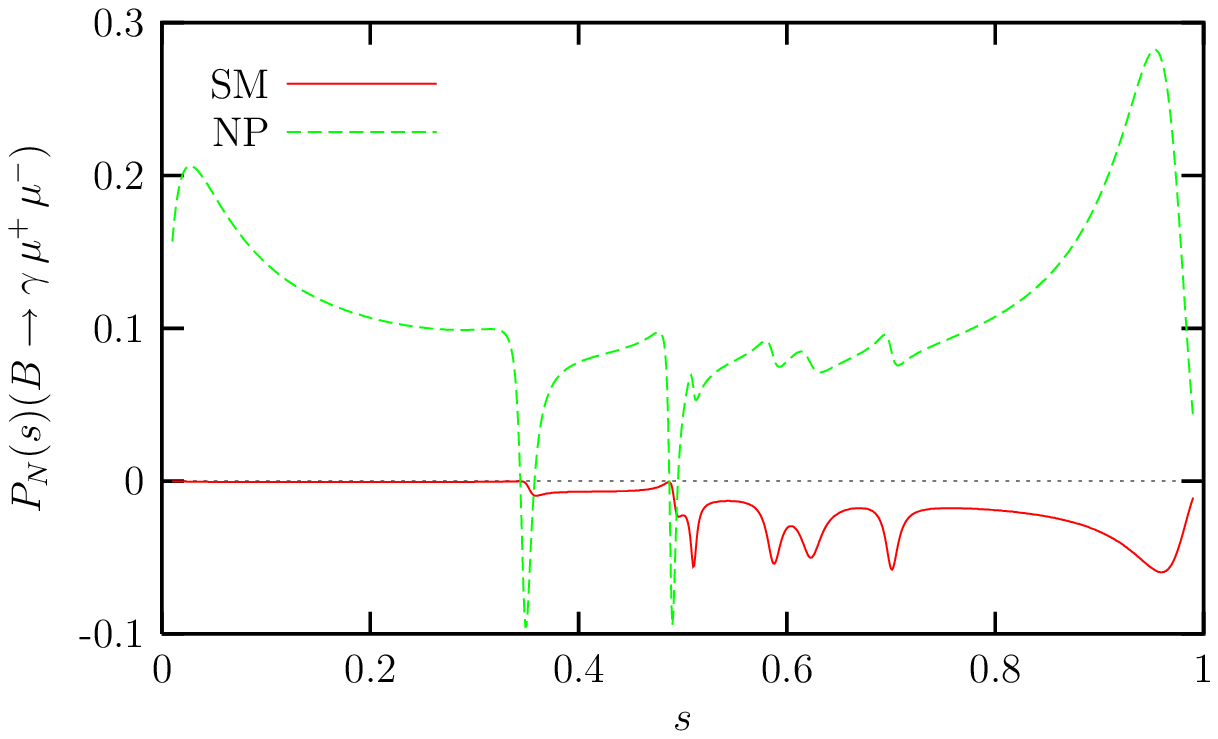}
\caption{The dependence of the  normal polarization
asymmetry $P_N (s)$ of $\ell^-$  for the $B_s \rar \gamma \, \mu^+ \mu^-$
decay on $s$ \label{f3}.}
\end{figure}

\begin{figure}
\centering
\includegraphics[width=5in]{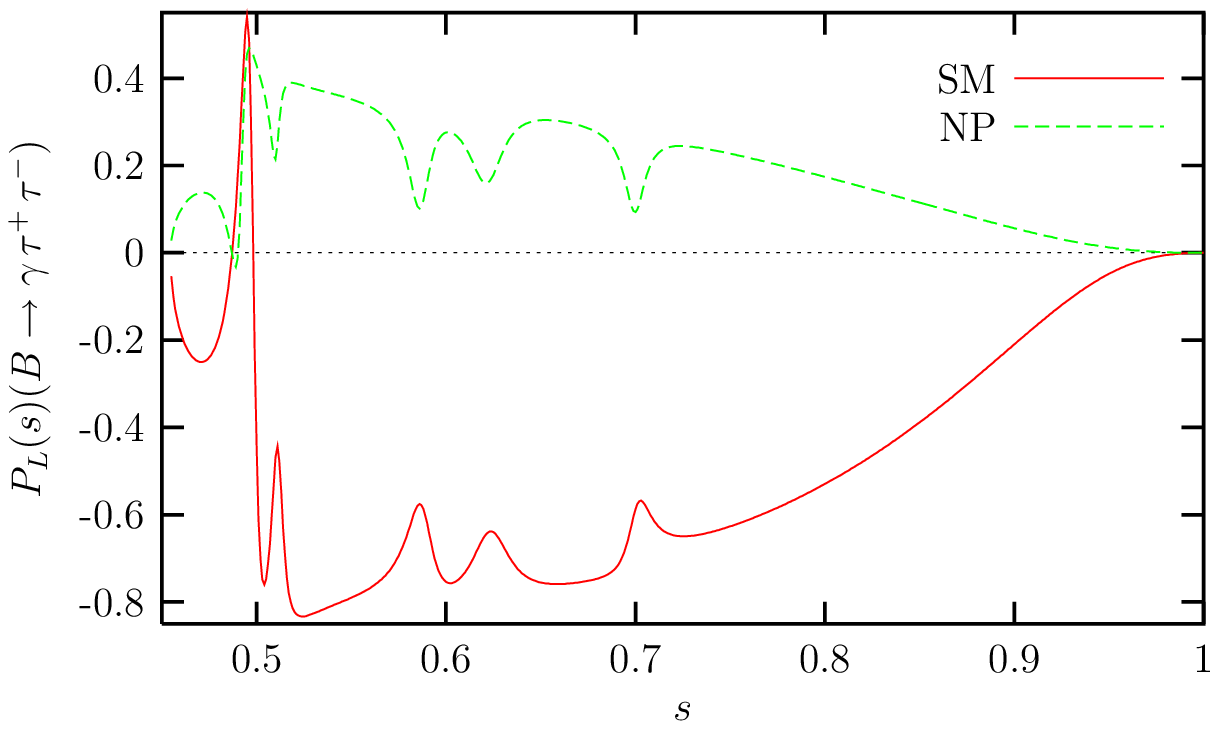}
\caption{The same as Fig.(\ref{f1}), but for the $B_s \rar \gamma
\, \tau^+ \tau^-$  decay.\label{f4}}
\end{figure}

\begin{figure}
\centering
\includegraphics[width=5in]{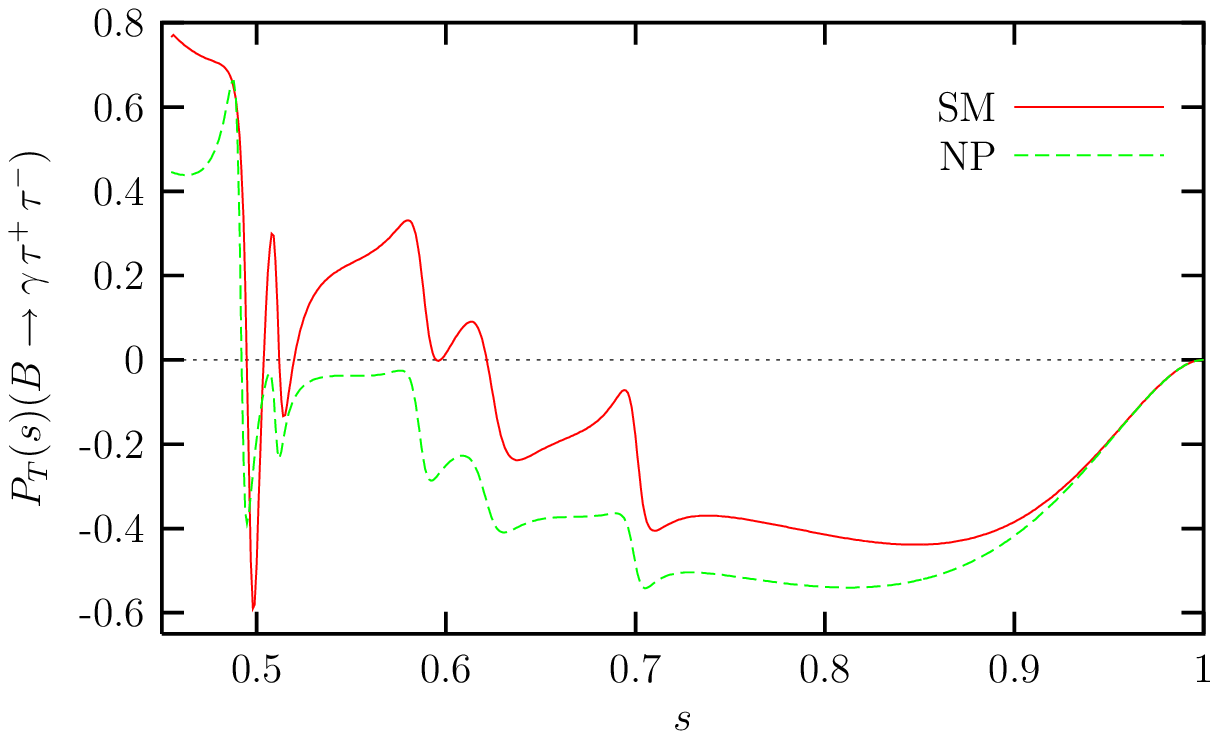}
\caption{The same as Fig.(\ref{f2}), but for the $B_s \rar \gamma
\, \tau^+ \tau^-$  decay \label{f5}.}
\end{figure}

\begin{figure}
\centering
\includegraphics[width=5in]{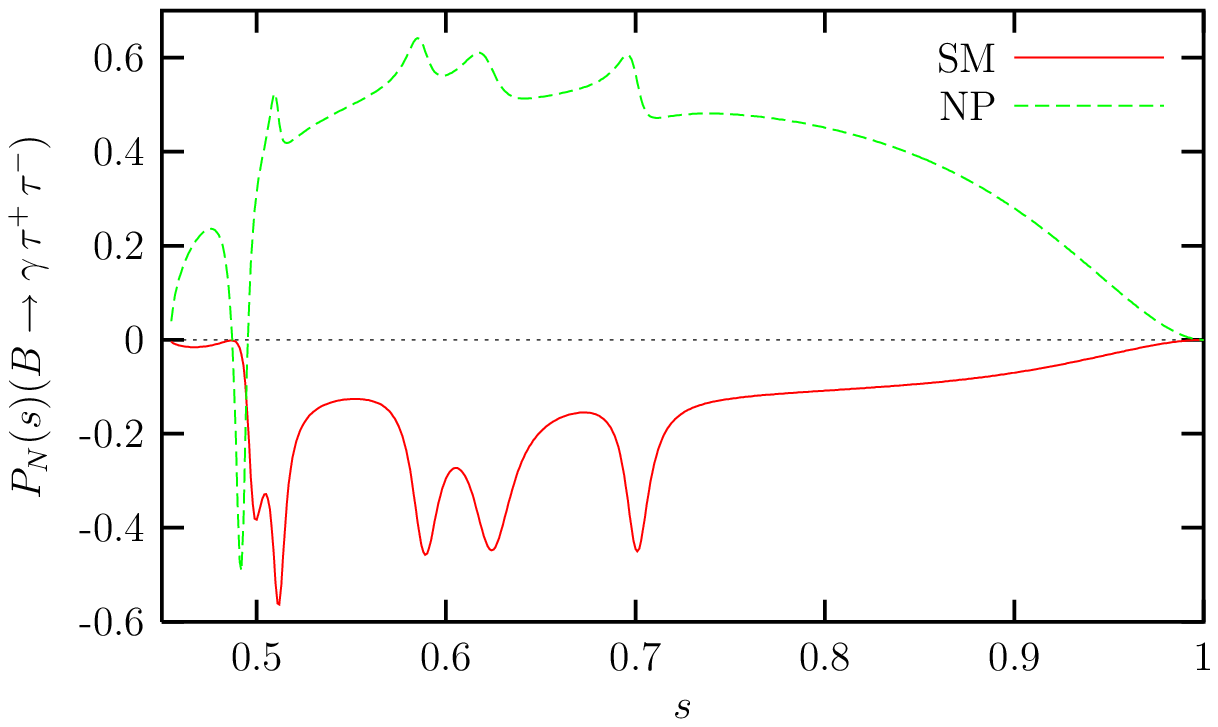}
\caption{The same as Fig.(\ref{f3}), but for the $B_s \rar \gamma
\, \tau^+ \tau^-$  decay.\label{f6}}
\end{figure}

In Figs. (\ref{f1})-(\ref{f6}) we present our results for the
various differential  polarization asymmetries  within the SM and also with
the new value of the coefficient $C_{10}$ given by (\ref{C10New}), which results
from the enhanced values of the Z penguins. In table (\ref{table2}), we
have given the averaged values of these asymmetries.
As  can be seen, the new value of $C_{10}$ can give substantial
changes in the SM results.
We note that due to the complex and enhanced value of the $bsZ$ vertex, the value of longitudinal
polarizations decrease   as compared to their SM
values for both decay modes, however, the transverse and  normal asymmetries show a substantial
increase
from their respective SM values. This increase is especially manifest for  $P_N$,
which changes its sign too as compared to its SM value
and   its magnitude increases by one order of magnitude for $\mu$ mode, and
by almost 200 $\%$ for $\tau$ mode.

Therefore,  future measurements of the enhanced normal and transverse
polarization asymmetry would be a suitable testing ground
for the validity of a complex $bsZ$ vertex and the model in ref. \cite{Buras1}.
\begin{table}[h]
\begin{tabular}{l | c c  c  c || c c c c }\hline &&&&& &&& \\[-0.3cm]Decay Mode
& \multicolumn{4}{|c||}{ $B_s \rar \gamma \, \mu^+ \mu^- $} &
\multicolumn{4}{|c}{ $B_s \rar \gamma \, \tau^+ \tau^- $} \\ &&&&& &&& \\[-0.3cm] \hline
&  BR $\times 10^8$ & ${\cal P}^-_L$ & ${\cal P}^-_T$ &
${\cal P}^-_N$ &  BR $\times 10^8$ & ${\cal P}^-_L$ & ${\cal P}^-_T$ &
${\cal P}^-_N$ \\ \hline
SM & 1.51 & - 0.85  &   - 0.07  &  - 0.01  & 1.14 & - 0.23  &   - 0.19  &  - 0.07   \\
enhanced bsZ & 3.70 & 0.08  &   -0.15  &  0.10  & 4.68 & 0.07  &   -0.25  &  0.22  \\ \hline
\end{tabular}
\caption{Predictions of the observables.} \label{table2}
\end{table}

%
\newpage
\renewcommand{\topfraction}{.99}
\renewcommand{\bottomfraction}{.99}
\renewcommand{\textfraction}{.01}
\renewcommand{\floatpagefraction}{.99}

\clearpage
\clearpage
\end{document}